\documentclass[12pt]{article}

\usepackage{amsmath,amssymb,graphicx,color,array,cite}
\numberwithin{equation}{section}

\evensidemargin \oddsidemargin

\begin{document}

\newcommand{\story}{\vspace{5mm} \noindent $\spadesuit$ }

\begin{titlepage}

\renewcommand{\thefootnote}{\fnsymbol{footnote}}


\begin{flushright}
CQUeST-2007-0081 \\
arXiv:yymm.nnnn
\end{flushright}

\vspace{15mm}
\baselineskip 9mm
\begin{center}
  {\Large \bf Hawking Radiation from Non-Extremal D1-D5 \\
  Black Hole via Anomalies}
\end{center}

\baselineskip 6mm
\vspace{10mm}
\begin{center}
  Hyeonjoon Shin\footnote{\tt hshin@sogang.ac.kr} 
  \\[3mm] 
  {\sl Center for Quantum Spacetime, Sogang University, Seoul
    121-742, South Korea }
  \\[10mm]
  Wontae Kim\footnote{\tt wtkim@sogang.ac.kr}
  \\[3mm]
  {\sl Department of Physics and Center for Quantum Spacetime \\
    Sogang University, C.P.O. Box 1142, Seoul 100-611, South Korea}
  \\[3mm]
\end{center}

\thispagestyle{empty}

\vfill
\begin{center}
{\bf Abstract}
\end{center}
\noindent
We take the method of anomaly cancellation for the derivation of
Hawking radiation initiated by Robinson and Wilczek, and apply it to
the non-extremal five-dimensional D1-D5 black hole in string theory.
The fluxes of the electric charge flow and the energy-momentum tensor
from the black hole are obtained. They are shown to match exactly with
those of the two-dimensional black body radiation at the Hawking
temperature.
\\ [5mm]

\vspace{5mm}
\end{titlepage}

\baselineskip 6.6mm
\renewcommand{\thefootnote}{\arabic{footnote}}
\setcounter{footnote}{0}

\section{Introduction}

Hawking radiation first observed by Hawking \cite{hawking} is the
quantum effect of fields in a classical space-time background with an
event horizon.  Although it is semi-classical, it provides an
important key to understand the nature of black hole horizon.  Since
the quantum effect of gravity itself becomes no longer negligible near
the black hole horizon, Hawking radiation also provides a basic
information in formulating the theory of quantum gravity, like the
string theory which is one of the candidates.

Usually, a certain physical result may have various mathematical
formulations for obtaining it and interpretations from various
different angles.  Having various viewpoints is always useful and
important in deepening the understanding of it.  As for Hawking
radiation, a new interpretation has been proposed by Robinson and
Wilczek \cite{rw}.  They have shown that the Hawking radiation plays
the role of preserving general covariance at the quantum level by
canceling the diffeomorphism anomaly at the event horizon.  Actually,
there is a similar work \cite{cf} that also considers the Hawking
radiation from the viewpoint of anomaly.  However, as noted in
\cite{rw}, it is specialized to two-dimensional space-time.  On the
other hand, the derivation of Hawking radiation based on anomaly
cancellation at the horizon does not depend on the space-time
dimension, and confirms that Hawking radiation is a universal
phenomenon.

The proposal by Robinson and Wilczek, which is based on the static and
spherically symmetric black hole, has been elaborated in
\cite{iso1,iso2} where, via extensions to charged and rotating black
holes, it has been shown that Hawking radiation is capable of
canceling anomalies of local symmetries at the horizon.  After this
elaboration, there have been many subsequent works which apply the
method of anomaly cancellation to various black holes in various
dimensions and verify the validity of the method
\cite{murata}-\cite{kui}. Further investigation on the derivation of
Hawking flux itself has been also given in \cite{iso4}.  In this
paper, we give one more example supporting and confirming the method
of anomaly cancellation by considering a typical black hole background
in string theory.  It is expected that our result strengthens the
validity and power of the method.

The black hole background we are concerned about is the charged
non-extremal five-dimensional black hole in string theory, which is
obtained from a specific D-brane configuration and often called the
non-extremal D1-D5 black hole \cite{hms}.\footnote{For detailed and
  comprehensive review on non-extremal D1-D5 black hole, see for
  example \cite{rev:d1-d5}.}  This background is particularly
interesting since, as noted in \cite{hms}, it is related to various
black solutions by taking different limits on parameters appearing in
the background; five-dimensional Reissner-Nordstr\"{o}m and
Schwarzschild solutions, six-dimensional black string solution
\cite{hhs}, black five-brane solution \cite{hs}, dyonic black string
solution \cite{hs2}.  So it may be argued that Hawking radiation from
several black backgrounds can be discussed by considering just one
background.

The organization of this paper is as follows: After a brief
description on the non-extremal five-dimensional D1-D5 black hole from
the in the next section, we consider a test charged scalar field in
the black hole background in Sec.~\ref{qf}, and show that, near the
horizon, the action for the scalar field reduces to a two-dimensional
theory in a certain background.  In Sec.~\ref{fluxes}, we calculate
the fluxes of the electric charge flows and the energy-momentum tensor
by applying the method of anomaly cancellation to the effective
two-dimensional theory, and show that the results match exactly with
the fluxes of black body radiation at Hawking temperature.  Finally,
the discussion follows in Sec.~\ref{discuss}.

\section{Non-extremal five-dimensional D1-D5 black hole}
\label{bh}

The non-extremal five-dimensional black hole originates from a brane
configuration in Type IIB superstring theory compactified on $S^1
\times T^4$.  The configuration relevant to the present case is
composed of D1-branes wrapping $S^1$, D5-branes wrapping $S^1 \times
T^4$, and momentum modes along $S^1$.  The solution of the Type IIB
supergravity corresponding to this configuration is a supersymmetric
background known as the extremal five-dimensional D1-D5 black hole.
The extremal black hole preserves some fraction of supersymmetry and
hence has zero Hawking temperature, which implies that we do not see
Hawking radiation.  Therefore, in order to consider the Hawking
radiation, we need the non-extremal version of the extremal solution.

Let $x_5$ and $x_6,\dots, x_9$ be periodic coordinates along $S^1$ and
$T^4$, respectively.  Then the ten-dimensional supergravity background
corresponding to the non-extremal D1-D5 black hole has the following
form in the string frame \cite{hms}:
\begin{align}
ds^2_{10} &= f_1^{-1/2} f_5^{-1/2} 
             ( - h f_n^{-1} dt^2 
               + f_n ( dx_5 + (1-\tilde{f}_n^{-1}) dt)^2 )
\notag \\
&+ f_1^{1/2} f_5^{-1/2} (dx_6^2+ \cdots + dx_9^2)
+ f_1^{1/2} f_5^{1/2} ( h^{-1} dr^2 + r^2 d \Omega_3^2) ~, 
\notag \\
e^{-2 \phi} &= f_1^{-1} f_5 ~, \quad
C_{05} = \tilde{f}_1^{-1} -1  ~,
\notag \\
F_{ijk} &= \frac{1}{2} \epsilon_{ijkl} 
          \partial_l \tilde{f}_5~,\quad
i,j,k,l=1,2,3,4~,
\label{10dsol}
\end{align}
where $F$ is the three-form field strenth of the RR 2-form gauge
potential $C$, $F = dC$. Various functions appearing in the background
are functions of coordinates $x_1, \dots, x_4$ given by
\begin{gather}
h = 1 - \frac{r_0^2}{r^2} ~, \quad
f_{1,5,n} = 1+ \frac{r_{1,5,n}^2}{r^2} ~, 
\notag \\
\tilde{f}_{1,n}^{-1} 
    = 1 - \frac{r_0^2 \sinh \alpha_{1,n} 
                 \cosh \alpha_{1,n}}{r^2} f_{1,n}^{-1} ~,
\notag \\
r^2_{1,5,n} = r_0^2 \sinh^2 \alpha_{1,5,n} ~, \quad
r^2 = x_1^2 + \cdots + x_4^2~,
\end{gather}
where $r_0$ is the extremality parameter. Here, $h$ and $f_{1,5,n}$,
are harmonic functions representing the non-extremality and the
presence of D1, D5, and momentum modes, respectively.

Upon dimensional reduction of Eq.~(\ref{10dsol}) along $S^1 \times
T^4$ following the procedure of \cite{mss}, we get the Einstein metric
of the non-extremal five-dimensional black hole as
\begin{align}
ds^2_5 = -\lambda^{-2/3} h dt^2 + 
\lambda^{1/3} (h^{-1} dr^2 + r^2 d \Omega_3^2) ~,
\label{5dmetric}
\end{align}
where $\lambda$ is defined by
\begin{gather}
\lambda = f_1 f_5 f_n ~.
\end{gather}
The location of the event horizon, $r_H$, of this black hole geometry
is obtained as
\begin{align}
r_H = r_0 ~.
\end{align}

Apart from the metric, the dimensional reduction gives us three kinds
of gauge fields.  The first one is the Kaluza-Klein gauge field
$A^{(K)}_\mu$ coming from the metric, and the second one, say
$A^{(1)}_\mu$, basically stems from $C_{\mu 5}$.  (We note that $\mu =
0,1,2,3,4$.) From the background of Eq.~(\ref{10dsol}), two gauge
fields are obtained as
\begin{align}
A^{(K)} = -(\tilde{f}_n^{-1} - 1 )dt ~, \quad
A^{(1)} = ( \tilde{f}_1^{-1} - 1 )dt ~.
\label{gauge}
\end{align}
Unlike these gauge fields which are one-form in nature, the last one
is the two-form gauge field, $A_{\mu\nu}$, originating from
$C_{\mu\nu}$, whose field strength is given by the expression of $F$
in Eq.~(\ref{10dsol}).  Though this two-form gauge field gives a
non-zero contribution to the full black hole background, it will not
play any role in the remaining part of this paper, and thus be
excluded in our consideration from now on.  Then, the background
composed of Eqs.~(\ref{5dmetric}) and (\ref{gauge}) will be our
concern.

\section{Quantum field near the horizon}
\label{qf}

In this section, we consider a free complex scalar field in the black
hole background, Eqs.~(\ref{5dmetric}) and (\ref{gauge}), and
investigate its action near the horizon based on the observation of
Ref.~\cite{rw}.  The field is taken to have minimal coupling to the
gauge fields, Eq.~(\ref{gauge}).  We would like to note that this
gives a simple reason why the two-form gauge field does not enter
seriously in our study; the object minimally coupled to the two-form
gauge field is not point-like but string-like one.

The action for the complex scalar field $\varphi$ in the background,
Eqs.~(\ref{5dmetric}) and (\ref{gauge}), is evaluated as
\begin{align}
S[\varphi]
&= - \int d^5 x \sqrt{-g} g^{\mu\nu} 
             (D_\mu \varphi)^* D_\nu \varphi 
\notag \\
&= - \int dt dr  \, r^3 \!\! \int d \Omega_3 \, \varphi^*
\bigg( 
	- \frac{\lambda}{h} D_t^2 
	+\frac{1}{r^3} \partial_r r^3 h \partial_r
	+\frac{1}{r^2} \nabla^2_\Omega
\bigg) \varphi ~,
\label{saction}
\end{align}
where $\int d\Omega_3$ and $\nabla^2_\Omega$ denote the integration
and the Laplacian on unit three sphere, respectively, and $D_t
= \partial_t - i e_1 A^{(1)}_t - i e_K A^{(K)}_t$ with the $U(1)$
charges $e_1$ and $e_K$ is the covariant derivative.

First of all, we perform the partial wave decomposition of $\varphi$
in terms of the spherical harmonics on $S^3$ as $\varphi = \sum_a
\varphi_a Y_a$, where $a$ is the collection of angular quantum numbers
of the spherical harmonics and $\varphi_a$ depends on the coordinates,
$t$ and $r$.  Then we see that the action is reduced to a
two-dimensional effective theory with an infinite collection of fields
labeled by $a$.  Next, in order to see what happens near the horizon,
it is helpful to take a transformation to the tortoise coordinate
$r^*$, which, in our case, is defined by
\begin{align}
\frac{\partial r^*}{\partial r} =
\frac{\lambda^{1/2}}{h} \equiv \frac{1}{f(r)} ~, 
\end{align}
and leads to $\int dr = \int dr^* f(r(r^*))$.  In the region near the
horizon, $f(r(r^*))$ (or $h(r(r^*))$) appears to be a suppression
factor vanishing exponentially fast, and thus the terms in the action
which do not have some factor compensating it can be ignored.  In our
case, the terms coming from the Laplacian on unit three sphere are
suppressed by $f(r(r^*))$.  We note that the suppression also takes
place for the mass term or the interaction terms of $\varphi$ when
they are included in the action (\ref{saction}).  Therefore, quite
generically, the action near the horizon becomes
\begin{align}
S[\varphi] = - \sum_a \int dt dr r^3 \lambda^{1/2}
  \varphi^*_a 
  \left( - \frac{1}{f} (\partial_t -i A_t)^2 
          + \partial_r f \partial_r 
  \right)
  \varphi_a ~,
\label{action}
\end{align}
where $A_t = e_1 A^{(1)}_t + e_K A^{(K)}_t$.  Now it is not hard to
find that this action describes an infinite set of massless
two-dimensional complex scalar fields in the following background:
\begin{gather}
ds^2 = - f(r) dt^2 + \frac{1}{f(r)} dr^2~,\quad
\Phi = r^3 \lambda^{1/2} ~,\notag \\
A_t = - \frac{e_1 r_0^2 \sinh \alpha_1 \cosh \alpha_1}{r^2 + r_1^2}
    + \frac{e_K r_0^2 \sinh \alpha_n \cosh \alpha_n}{r^2 + r_n^2}~,
\label{2dbg}
\end{gather}
where $\Phi$ is the two-dimensional dilaton field.

What we have seen is that the physics near the horizon of the original
five-dimensional theory (\ref{saction}) is effectively described by a
two-dimensional theory, which is non-interacting and massless one
(\ref{action}).

\section{Anomalies and Hawking fluxes}
\label{fluxes}

Having the two-dimensional effective field theory near the horizon
(\ref{action}), we consider the problem of Hawking radiation following
the approach based on the anomaly cancellation proposed in
\cite{rw,iso1}.

One important ingredient of the anomaly approach of \cite{rw} is to
notice that, since the horizon is a null hypersurface, all ingoing
(left moving) modes at the horizon can not classically affect physics
outside the horizon.  This implies that they may be taken to be out of
concern at the classical level and thus the effective two-dimensional
theory becomes chiral, that is, the theory only of outgoing (right
moving) modes.  If we now perform the path integration of right moving
modes, the resulting quantum effective action becomes anomalous under
the gauge or the general coordinate transformation, due to the absence
of the left moving modes. However, such anomalous behaviors are in
contradiction to the fact that the underlying theory is not anomalous.
The reason for this is simply that we have ignored the quantum effects
of the classically irrelevant left moving modes at the horizon.  Thus
anomalies must be cancelled by including them.  In what follows,
anomaly cancellations at the horizon are studied and their relation to
the Hawking fluxes is investigated.

The previous paragraph states that anomalies appear at the horizon
$r_H$.  For computational convenience, we regard the quantum effective
action to be anomalous in an infinitesimal slab, $r_H \le r \le r_H +
\epsilon$, which is the region near the horizon.  (The limit $\epsilon
\rightarrow 0$ is taken at the end of the calculation.)  This leads to
a splitting of the region outside the horizon, $r_H \le r \le \infty$,
into two regions, $r_H \le r \le r_H + \epsilon$ and $r_H + \epsilon
\le r \le \infty$.  Then, since the field we are considering is
charged one, there will be the gauge and the gravitational anomaly
near the horizon, $r_H \le r \le r_H + \epsilon$.

We first consider the gauge anomaly.  Since there are two kinds of
$U(1)$ gauge symmetries, we have two $U(1)$ gauge currents, which are
denoted as $J^{(1)}_\mu$ and $J^{(K)}_\mu$ following the notation of
the original gauge potentials $A^{(1)}_\mu$ and $A^{(K)}_\mu$.  The
two-dimensional anomalies for these two current are identical in
structure.  So we will concentrate on the anomaly for $J^{(1)}_\mu$
and give just the result for another current.

Since the region outside the horizon has been divided into two
regions, it is natural to write the gauge current as a sum
\begin{align}
J^{(1)\mu} = J_{(o)}^{(1)\mu} \Theta_+(r) 
          +  J_{(H)}^{(1)\mu} H(r) ~,
\label{jsplit}         
\end{align}
where $\Theta_+(r) = \Theta(r-r_+-\epsilon)$ and $H(r)=1-\Theta_+(r)$.
Apart from the near horizon region, the current is conserved
\begin{align}
\partial_r J_{(o)}^{(1)r} =0 ~.
\label{joeq}
\end{align}
On the other hand, the current near the horizon is anomalous and obeys
the anomalous equation
\begin{align}
\partial_r J_{(H)}^{(1)r} = \frac{e_1}{4 \pi} \partial_r A_t ~,
\label{jheq}
\end{align}
which is the form of two-dimensional consistent gauge anomaly
\cite{bertlmann,bz}.  Since these two equations in each region are
first order differential ones, they can be easily integrated as
\begin{align}
J_{(o)}^{(1)r} &= c^{(1)}_o, \notag \\
J_{(H)}^{(1)r} &= c^{(1)}_H 
   + \frac{e_1}{4\pi} \left( A_t(r) -A_t(r_H) \right),
\label{jsol}
\end{align}
where $c^{(1)}_o$ and $c^{(1)}_H$ are integration constants.  We note
that $c^{(1)}_o$ is the electric charge flux which we are going to
obtain.

Now, we let $W$ be the quantum effective action of the theory without
including the ingoing (left moving) modes near the horizon.  Then its
variation under a gauge transformation with gauge parameter $\zeta$ is
given by
\begin{align} 
-\delta W 
& =\int d^2 x \sqrt{-g} \; \zeta \nabla_{\mu} J^{(1)\mu}
\notag \\
&= \int d^2 x \; \zeta 
\left[
      \partial_r \left( \frac{e_1}{4\pi}A_t H \right)  
    + \delta(r-r_+ - \epsilon) 
      \left( J_{(o)}^{(1)r} - J_{(H)}^{(1)r} 
             + \frac{e_1}{4 \pi}A_t \right)
\right] ~,
\label{gvar}
\end{align} 
where Eqs.~(\ref{jsplit}), (\ref{joeq}), and (\ref{jheq}) have been
used for obtaining the second line.  As alluded to in the early part
of this section, the full quantum effective action of the underlying
theory must have gauge invariance.  The full effective action includes
the quantum effects of the ingoing modes near the horizon, whose gauge
variation gives a term canceling the first term of (\ref{gvar}).  For
the gauge invariance, the coefficient of the delta function in
Eq.~(\ref{gvar}) should also vanish, and hence, by using
Eq.~(\ref{jsol}), we get
\begin{align}
c^{(1)}_o = c^{(1)}_H - \frac{e_1}{4\pi} A_t(r_H) ~.
\end{align}
In order to determine the charge flux $c^{(1)}_o$, the value of the
current at the horizon, $c^{(1)}_H$, should be fixed.  This is done by
imposing a condition that the covariant current \cite{bz} given by
$\tilde{J}^{(1)r} = J^{(1)r} + \frac{e_1}{4\pi} A_t(r) H(r)$ vanishes
at the horizon, which, as noted in \cite{iso2}, assures the regularity
of physical quantities at the future horizon. Then, the electric
charge flux canceling gauge anomaly is determined as
\begin{align}
c^{(1)}_o =  - \frac{e_1}{2 \pi} A_t(r_H) 
= \frac{e_1}{2\pi} ( e_1 \tanh \alpha_1 - e_K \tanh \alpha_n) ~.
\label{co1}
\end{align}

As for the current $J^{(K)}_\mu$ associated with another $U(1)$ gauge
symmetry, we can follow the same steps from Eq.~(\ref{jsplit}) to
Eq.~(\ref{co1}), with the anomaly equation
\begin{align}
\partial_r J_{(H)}^{(K)r} = \frac{e_K}{4 \pi} \partial_r A_t ~,
\label{jhkeq}
\end{align}
and obtain
\begin{align}
c^{(K)}_o = - \frac{e_K}{2 \pi} A_t(r_H) 
= \frac{e_K}{2\pi} ( e_1 \tanh \alpha_1 - e_K \tanh \alpha_n) ~.
\label{coK}
\end{align}

As we will see, the electric charge fluxes, (\ref{co1}) and
(\ref{coK}), exactly match with those of the two-dimensional Hawking
(blackbody) radiation with the Planck distribution including chemical
potentials.

We now turn to the problem of determining the flux of the
energy-momentum tensor through the cancellation of the gravitational
anomaly.  The method for solving it is the same with that adopted in
the case of gauge anomaly.  First of all, like the splitting of
Eq.~(\ref{jsplit}), we write the energy-momentum tensor as
\begin{align}
T^\mu_\nu = T^\mu_{\nu(o)}\Theta_+(r) +  T^\mu_{\nu(H)} H(r) ~.
\label{tsplit}         
\end{align}
Due to the presence of the gauge potentials and the dilaton in the
background (\ref{2dbg}), the energy-momentum tensor satisfies the
modified conservation equation \cite{iso1}.  What is of interest for
our problem is the conservation equation for the component $T^r_t$,
the energy-momentum flux in the radial direction.  Apart from the near
horizon region, it is given by
\begin{align}
\partial_r T^r_{t(o)} = J^r_{(o)} \partial_r A_t ~.
\label{toeq}
\end{align}
Here $J^r_{(o)}$ comes from the current $J^r \equiv \frac{1}{e_1}
J^{(1)r} = \frac{1}{e_K} J^{(K)r}$ in a splitting like
Eq.~(\ref{jsplit}) and satisfies $\partial_r J^r_{(o)}=0$, whose
solution is $J^r_{(0)} = c_0$ with $c_0 = \frac{1}{e_1} c^{(1)}_o$ or
$\frac{1}{e_K} c^{(K)}_o$.  In the near horizon region, we have
anomalous conservation equation \cite{iso1} as
\begin{align}
\partial_r T^r_{t(H)} 
=  J_{(H)}^r \partial_r A_t + A_t \partial_r J_{(H)}^r
 + \partial_r N^r_t ~,
\label{theq}
\end{align}
where $ N^r_t =( f^{\prime 2}+f f^{\prime\prime})/192\pi$.  (The prime
denotes the derivative with respect to $r$.)  The second term comes
from the gauge anomaly represented by the anomalous conservation
equation $\partial_r J^r_{(H)} = \frac{1}{4\pi} \partial_r A_t$, while
the third term is due to the gravitational anomaly for the consistent
energy-momentum tensor \cite{aw}.  Now it is not a difficult task to
integrate Eqs.~(\ref{toeq}) and (\ref{theq}) and obtain
\begin{align}
T^r_{t(o)} &= a_o + c_o  A_t ~, \notag \\
T^r_{t{(H)}} &= a_H + \int^r_{r_H} dr \partial_r 
    \left( c_o A_t + \frac{1}{4\pi}A_t^2 + N^r_t \right) ~,
\label{tsol}
\end{align}
where $a_o$ and $a_H$ are integration constants. Here $a_o$ is the
energy flux which we are interested in.

Next, we consider the variation of quantum effective action $W$ under
a general coordinate transformation in the time direction with a
transformation parameter $\xi^t$:
\begin{align}
- \delta W 
&= \int d^2x \sqrt{-g} \; \xi^t \nabla_\mu T^\mu_{t} 
\notag \\
&= \int d^2x \; \xi^t
  \bigg[ c_o \partial_r A_t +
        \partial_r 
        \left[ \left( \frac{1}{4\pi} A_t^2 + N^r_t \right) H
        \right]       
\notag \\
&+ \left( T^r_{t~(o)} - T^r_{t~(H)} 
         + \frac{1}{4\pi}A_t^2+N^r_t
   \right) \delta(r-r_+ -\epsilon) 
     \bigg] ~.
\end{align}
The first term in the second line is purely the classical effect of
the background electric field for constant current flow.  The second
term is cancelled by including the quantum effect of the ingoing modes
as is the case of gauge anomaly.  The last term gives non-vanishing
contribution at the horizon and is also required to vanish for the
general covariance of the full quantum effective action.  This
requirement leads us to have the following relation.
\begin{align}
a_o = a_H +\frac{1}{4\pi}A_t^2(r_H)- N^r_t(r_H) ~,
\end{align}
where the solution Eq.~(\ref{tsol}) has been used.  For determining
$a_o$, we first need to know the value of $a_H$, which is fixed by
imposing a condition that the covariant energy-momentum tensor
vanishes at the horizon for regularity at the future horizon
\cite{iso2}.  Then, from the expression of the covariant
energy-momentum tensor \cite{bz,bk}, $\tilde{T}^r_t = T^r_t
+\frac{1}{192\pi} (f f'' -2(f')^2)$, the condition
$\tilde{T}^r_t(r_H)=0$ gives
\begin{align}
a_H= \frac{\kappa^2}{24 \pi} = 2N^r_t(r_H) ~,
\end{align}
where $\kappa$ is the surface gravity at the horizon,
\begin{align}
\kappa = 2\pi T_H = \frac{1}{2} \partial_r f |_{r=r_H} =
\frac{1}{ r_0 \cosh \alpha_1 \cosh \alpha_5 \cosh \alpha_n}~.
\end{align}
Here we see that the Hawking temperature of the non-extremal D1-D5
black hole is
\begin{align}
T_H = 
\frac{1}{2 \pi r_0 \cosh \alpha_1 \cosh \alpha_5 \cosh \alpha_n}~,
\end{align}
which is the desired correct value.  Having the value of $a_H$, the
flux of the energy-momentum tensor is finally determined as
\begin{align}
a_o 
&= \frac{1}{4\pi}A_t^2(r_H) +N^r_t(r_+) \notag \\
&= \frac{1}{4\pi} (e_1 \tanh \alpha_1 - e_K \tanh \alpha_n)^2
 +\frac{\pi}{12} T_H^2 ~,
\label{tflux}
\end{align}
which matches exactly with that of the Hawking radiation from the
black hole as will be shown below.

Up to now, we have obtained the fluxes of electric charges,
Eqs.~(\ref{co1}) and (\ref{coK}), and energy-momentum tensor,
Eq.~(\ref{tflux}) via the method of anomaly cancellation.  It is an
interesting and important problem to check that these results coincide
with the usual fluxes of Hawking (black body) radiation from the black
hole.  Although the radiation in the case of bosons should be treated,
we simply consider the fermion case in order to avoid the
superradiance problem.  The Hawking distribution for fermions is given
by the Planck distribution at the Hawking temperature with two
electric chemical potentials for the charges $e_1$ and $e_K$ of the
fields radiated from the black hole,
\begin{align}
N_{e_1,e_K}(\omega) = 
\frac{1}{e^{(\omega - e_1 \Phi_1 -e_K \Phi_K)/T_H} +1} ~,
\end{align}
where $\Phi_1 = \tanh \alpha_1$ and $\Phi_K = \tanh \alpha_n$.  By
using this, the electric charge fluxes of Hawking radiation, say $F_1$
and $F_K$, can be calculated as
\begin{align}
F_1 &= e_1 \int^\infty_0 \frac{d\omega}{2\pi}
		( N_{e_1,e_K}(\omega) - N_{-e_1,-e_K}(\omega) ) 
\notag \\
	&= \frac{e_1}{2\pi} ( e_1 \tanh \alpha_1 - e_K \tanh \alpha_n)~,
\\
F_K &= e_K \int^\infty_0 \frac{d\omega}{2\pi}
		( N_{e_1,e_K}(\omega) - N_{-e_1,-e_K}(\omega) ) 
\notag \\
	&= \frac{e_K}{2\pi} ( e_1 \tanh \alpha_1 - e_K \tanh \alpha_n)~,	
\end{align}
which exactly match with Eqs.~(\ref{co1}) and (\ref{coK}). As for the
energy-momentum flux of Hawking radiation, say $F_E$, we can obtain
\begin{align}
F_E &= \int^\infty_0 \frac{d\omega}{2\pi}
		( N_{e_1,e_K}(\omega) + N_{-e_1,-e_K}(\omega) ) 
\notag \\
	&= \frac{1}{4\pi} (e_1 \tanh \alpha_1 - e_K \tanh \alpha_n)^2
       +\frac{\pi}{12} T_H^2~,	
\end{align}
which also shows the exact coincidence with the flux of
Eq.~(\ref{tflux}).  These exact matchings imply that, as first
realized in \cite{rw}, the fluxes of Hawking radiation from the black
hole we have been considered are capable of canceling the gauge and
the gravitational anomalies at the horizon.

\section{Discussion}
\label{discuss}

We have applied the method of anomaly cancellation for calculating the
Hawking radiation initiated by Robinson-Wilczek to the non-extremal
five-dimensional D1-D5 black hole in string theory, and obtained the
fluxes of the electric charge flow and the energy-momentum tensor. The
resulting fluxes match exactly with those of the two-dimensional black
body radiation at the Hawking temperature.  The point is that the
Hawking radiation plays the role of canceling possible gauge and
gravitational anomalies at the horizon to make the gauge and
diffeomorphism symmetry manifest at the horizon.  This confirms that
the anomaly analysis proposed in \cite{rw,iso1} is still working and
valid for a typical black hole in string theory.

What we have considered in the black hole background is the scalar
field, which corresponds to a point-like object, that is, point
particle.  As already mentioned, it cannot have minimal coupling to
the two-form gauge field.  This gives the basic reason that the
two-form gauge field does not enter the story.  One possibility for
introducing the effect of the two-form gauge field in the
two-dimensional action (\ref{action}) is to consider the dual gauge
field.  Note that the dual of the two-form gauge field in five
dimensions is one-form gauge field.  So, the field can couple
minimally to the dual gauge field, and the nature of the charge
carried by the field becomes magnetic from the viewpoint of the
original two-form gauge field.  It would be interesting to see what
one obtains when the dual field is also considered.

The present work is based purely on the viewpoint of quantum field
theory, though the black hole we are interested in has the string
theory origin.  In other words, we have not minded whether the complex
scalar field $\varphi$ is in the field contents of type IIB string
theory compactified on five torus.  Upon compactification, many moduli
fields appear in the low enegy supergravity action.  Some of them,
especially the fixed scalar, are distinguished from the usual scalar
field.  What we see when such fields are considered instead of the
field $\varphi$ in applying the method of anomaly cancellation may be
an interesting question.

\section*{Acknowledgments}
This work was supported by the Science Research Center Program of the
Korea Science and Engineering Foundation through the Center for
Quantum Spacetime (CQUeST) of Sogang University with grant number
R11-2005-021.  The work of H.S. was supported in part by grant
No.~R01-2004-000-10651-0 from the Basic Research Program of the Korea
Science and Engineering Foundation (KOSEF).

\end{document}